# pp-Solar Neutrino Spectroscopy: Return of the Indium Detector

R. S. Raghavan

*Bell Laboratories, Lucent Technologies, Murray Hill NJ 07974*
(Bell Labs Tech. Memo. 10009622-010606-19TM)

A new indium-loaded liquid scintillator (LS) with up to 15wt% In and high light output promises a breakthrough in the 25y old proposal for observing pp solar neutrinos ($\nu_e$) by *tagged* $\nu_e$ capture in $^{115}$In. Intense background from the natural β-decay of In, the single obstacle blocking this project till now, can be reduced by more than x100 with the new In-LS. Only *non-In* background remains, dramatically relaxing design criteria. Eight tons of In yields ~400 pp $\nu_e$/y after analysis cuts. With the lowest threshold yet, Q=118 keV, In is the most sensitive detector of the pp $\nu_e$ spectrum, the long sought touchstone for $\nu_e$ conversion.

PACS numbers:

The dominant part (>90%) of the neutrino ($\nu_e$) flux from the sun arises from the basic proton-proton (pp) reaction with a value practically independent of solar astrophysics. Direct detection of pp $\nu_e$ is thus of basic interest for particle physics because non-standard $\nu$'s can be probed by a well-characterized high-flux $\nu_e$ beam on an ultra-long baseline. A non-standard pp $\nu_e$ flux and spectral shape can identify and quantify $\nu_e$ conversion and its mechanism. However, high background endemic to the low energies of pp $\nu_e$ signals make direct detection by standard methods hopeless in practice.

In 1976 a new approach to solar $\nu_e$ detection--a *taggable* $\nu_e$ capture--was proposed with indium as the specific target.[1] $\nu_e$-Capture in $^{115}$In leads to an isomeric state (τ = 4.7 μs) in $^{115}$Sn, releasing a prompt electron-- the $\nu_e$ signal. Its energy directly measures the $\nu_e$ energy: $E_\nu = E_e + Q$. The low $\nu_e$ threshold Q=118 keV reaches most of the pp $\nu_e$ spectrum (0-420 keV). The signal electron can be tagged as the product of $\nu_e$ capture by a unique delayed space-time coincidence of radiations (116+497=613 keV) deexciting the isomeric state. With the ~96% abundance of $^{115}$In, the theoretical signal is ~365 pp $\nu_e$/yr in an attractively modest 4 ton mass of In[1,2].

In is the most sensitive probe yet for exploring pp $\nu_e$'s. For example, oscillatory shapes are predicted at lower energies of the pp spectrum by a new model of vacuum oscillations[3,4,5] ("Just-so2") with low $\Delta m^2 \sim 5 \times 10^{-12}$ eV$^2$ (Fig. 1). Among the known and proposed pp sensitive detectors, only In can observe the modulations clearly.

Despite these assets, the In idea is dormant because of a *single* obstacle: the natural β$^-$-decay of $^{115}$In (τ = 6.4x10$^{14}$y, β$_{max}$ = 495 keV). The intense β$^-$ background overlaps the In pp signal (0-302 keV) and indeed, severely hampers the operation of the $\nu_e$ tag. In the best In loaded liquid scintillators (In-LS) of the time[6], even the $^7$Be $\nu_e$ signal (744 keV) was hardly resolved from the In β spectral tail and even less so, the In $\nu_e$ tag (613 keV). In 1997 a new pp $\nu_e$ target, *stable* $^{176}$Yb[7] that presents no target decay problems was identified. Yb is now the target basis of the LENS project[8].

LS spectroscopy is ideal for massive low energy $\nu_e$ detectors if the basic problem of loading metal targets in the LS can be solved. A new technology for metal-LS developed at Bell Labs has led to *Yb*-LS that meets the rigid prescriptions of LENS[5]. The same method applied to In produced an *In*-LS much superior to those of ref. 6. The new In-LS offers, for the first time, the signal quality needed to reduce the In β *bremsstrahlung* (BS) background--the key obstacle till now-- by x100 or more. Only *non-In* background then remains, relaxing design criteria dramatically. Revisiting well-developed design concepts of the In detector[9,10] in the light of this progress, I conclude that the In project is technically feasible with broad safety margins.

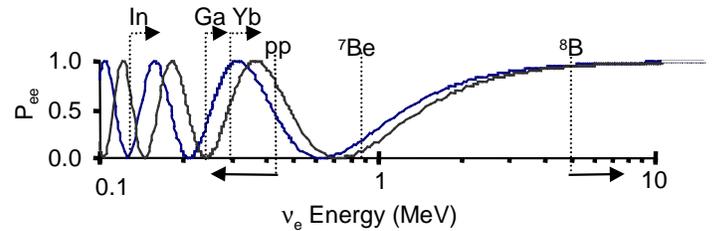

Fig. 1 $\nu_e$-Survival $P_{ee}$ vs. $\nu_e$ energy in the Just-so2 model for $\Delta m^2$ = 5.2 and 6x10$^{-12}$ eV$^2$.

A detector based on metal loaded LS such as Yb-LENS or In, demands extraordinary prescriptions: ~10 wt% target loading and a scintillation signal strength high enough for precision spectroscopy at very low energies <<100 keV in a 100-ton scale device with long term stability. Such a metal LS has never been produced till now.

The standard method for a metal-LS is dissolution of an organic salt of the metal in a high quality organic LS solvent. The procedure involves two basic aspects: 1) defining the lowest mass organic salt and compatible additives that can be dissolved in a LS solvent so that the final LS is practically free of scattering, absorption and chemical quenching of the scintillation light and 2) conversion of an inorganic salt of the metal into the selected organic salt and extraction into the LS solvent. While 2) is standard chemistry, 1) is far from predictable *a priori*. Thus, systematic empirical tests of a very large number of combinations of the salt, the solvent and various additives were carried out. The results set a roadmap for assembling the metal LS and optimizing it for a given target. The roadmap first led to a "neutrino" grade *Yb*-LS and its general applicability was proved by the production of the long sought high quality *In*-LS.

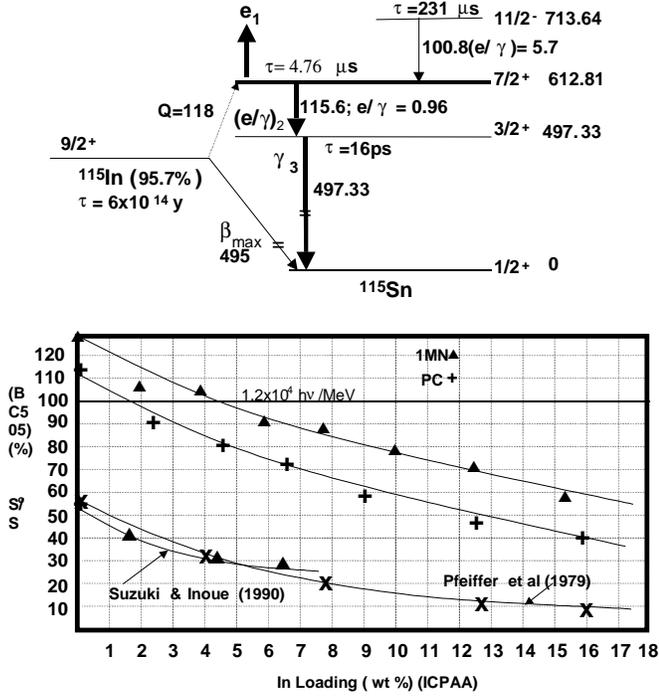

Fig. 2 (top) Level scheme of the $^{115}$In-$^{115}$Sn system; (bottom Scintillation data on the new Bell Labs In-LS (with 25 cc samples) vs. previous In-LS results.

Fig. 2 shows results for the new In-LS with two solvents, pseudocumene (PC) and 1-methylnaphthalene (MN). The scintillation yield relative to a LS standard, $S/[S_o=1.2 \times 10^4$ (h$\nu$)/ MeV], is plotted vs. the In loading (Fig.2). Compared to the previous best results, also shown, the new In-LS shows S values up to 3-5 times higher and the useful (i.e. S>50%) range of In-loading extended from <1% to 13- 16%. In ongoing studies of the optical transmission, a preliminary value of the 1/e transmission length of 9% In-LS(PC) at the practical wavelength of 430 nm, is ~2m. Work is in progress to improve this result. The S values and the chemical integrity of the samples have been stable over a few months so far. The preparative method is particularly simple and easily adaptable to large scale production.

The $^{115}$In-$^{115}$Sn $\nu_e$ capture reaction[1] (see Fig. 2) is:

$$\nu_e + {}^{115}In \rightarrow e^-_1 + {}^{115}Sn^* (613) \rightarrow (\text{Delay } \tau = 4.76 \text{ }\mu s) \rightarrow$$
$$(e/\gamma)(116) + \gamma_3 (497) + {}^{115}Sn \quad (1).$$

The $e_1$ electron spectrum leads directly to the $\nu_e$ spectrum.

*The Neutrino Tag*: Reaction (1) can be identified in extreme detail by a space-time coincidence $\nu_e$ tag "T" [1,6]:
(a) $e_1$–*delay*($\tau =4.76$ $\mu$s)- [$(e/\gamma)_2 + \gamma_3$] coincidence;
(b) $(e/\gamma)_2$-$\gamma_3$ in *prompt* coincidence (gate time ~10 ns);
(c) $e_1$-$(e/\gamma)_2$ in *spatial* coincidence in a primary "microcell" ;
(d) $\gamma_3$ shower triggers in at least 2 $\mu$-cells;
(e) $\gamma_3$ *contained* in "macrocell" surrounding primary $\mu$-cell;
(f) E$(e/\gamma)_2$ = 50-200 keV; E($\gamma_3$)= 450-750 keV;
(g) E[$\gamma_2+\gamma_3$] = 500-750 keV

*Measurement of Signal & Background:* The time spectrum T(a) contains $\nu_e$ signals in the exponential fall at "early" delays <10 $\mu$s and background in the nearly flat "late" delays of 10-200 $\mu$s. The background under the signal can thus be precisely *measured in vivo*, thus, a signal/noise s/n~1 is a necessary and sufficient design goal.

*The In decay Problem:* Consider a "standard" In detector with In mass I= $4 \times 10^6$g. The theoretical pp $\nu_e$ signal is S= $1/d/I = 1.16 \times 10^{-5}$/s. With a tag efficiency ~0.65$\varepsilon$ (see below) and time-window of $2\tau$ ~10$\mu$s (0.86 of the signal) the practical signal rate in I is S=6.5$\varepsilon$ $\times 10^{-6}$/s. The specific activity of In is 0.25Bq/g, thus the In singles rate in I is N= $10^6$/s. Reconciling the enormous disparity between N and S is the task of the tag T. To fulfill this task, the basic design feature needed is *granularity*. The number of $\mu$-cells $\eta$ is the parameter of final defence against the background N. The $\eta_{min}$(s/n =1) varies with background type.

*Topology of the $\nu_e$ Tag :* The energies of $(e/\gamma)_2$ and $\gamma_3$ in the In-Sn system are remarkably tailored for efficient spatial separation and containment of the tag components in the $\mu$- and macro-cells. A macrocell of radius 30 cm with 5-10% In fully contains ~95% of the 497 keV $\gamma_3$ triggers. Condition T(d) implies rejection of photo events of $\gamma_3$ in In (single triggers) that entails a signal loss <1to 1.7% for 5-10% In. In the $(e/\gamma)_2$ events, 45% are well localized conversion electrons $e_2$ and 55% are ~116 keV $\gamma_2$'s. Condition T(f) requires >50 keV deposit from $\gamma_2$ in the $\mu$-cell. Some 50% of $\gamma_2$ totally escape a typical $\mu$-cell. The $(e/\gamma)_2$ containment is thus (45+ 27) ~70% and the overall tag containment ≥65%. The detection efficiency $\varepsilon$ depends further on the cuts T(f,g).

*Uncorrelated random coincidences*: A candidate event satisfying the tag T can arise from 4 random In events via: i) 2 pulses in the same $\mu$-cell (In mass m) with a tag time of 10$\mu$s, ii) 2 pulses in 2 *different* $\mu$-cells in the macrocell (In mass M/m=500) within 10ns (mimicking $\gamma_3$); and iii) a coincidence of ii) with the *second* event in i) within 10ns. This quadruple random coincidence rate (measureable at late delays) can be compared to the signal S for s/n =1:

$R_4$ =$[(10^6/\eta)^2 10^{-5} \eta][(10^6/\eta)^2 (M/m)10^{-8}] \times 10^{-8}$ /s (2)
   =6.5$\varepsilon$ x $10^{-6}$/s
$\eta_{min}(4)$ ~4.2x$10^3$ (for M/m=500, $\varepsilon$ ~1) (3)

The $\eta(4)$ is the basic *irreducible* design granularity. Eq. (3) holds for all sources of uncorrelated, localized radiation such as pure $\alpha$, $\beta$'s from traces of U/Th ($\alpha$), $^{40}$K ($\beta$) and $^{14}$C ($\beta$). Their event rates, however, are much lower than that of In. The $^{14}$C($\beta$) activity e.g., is 0.2Bq/g of *modern* C (cf. 0.25 Bq/g In). Most LS organics contain ~$10^5$ times less $^{14}$C.Thus, $^{14}$C, normally the impenetrable barrier for pp $\nu_e$ detection, is not an issue for an LS-based In detector.

*Correlated chance coincidences-Bremsstrahlung (BS):* Instead of 3 random events mimicking $(e/\gamma)_2+\gamma_3$ in the tag, ($\beta$, $\gamma$) cascades can directly mimic $(e/\gamma)_2+ \gamma_3$. Though In is a pure $\beta$-emitter, the $\beta$'s radiate BS which can ideally mimic $(e/\gamma)_2+ \gamma_3$ because an In $\beta$ can deposit ~50 keV or more of its kinetic energy in a μ-cell and radiate the rest into the macro-cell, inseparable from the tag. Even though such events at the tail of the BS-spectrum are rare, the high In β rate makes it the most dangerous background. The two prompt coincidence conditions in eq. (2) do not apply since the event is correlated. The BS problem is thus more difficult than the β emission. The only defence is discrimination of the $\nu_e$ tag *energy* of 616 keV from the 495 keV BS endpoint. It is this aspect that sets a high premium on the LS *energy resolution*.

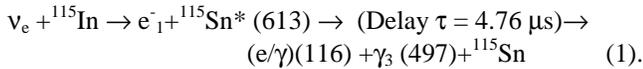

The differential BS energy spectrum of the In β-decay in thick-target In metal has been calculated[11] for ΔT = 50 keV, i.e. radiative transitions retaining 50 keV β kinetic energy (as required in T(f)). This BS spectrum was folded-in with the detector resolution $\sigma = k\sqrt{E(keV)}$ with k= 2.86, 2.24, 1.83 and 1.41 for the luminosities L=123, 200, 300, 500 photoelectrons (pe)/MeV respectively. The dashed curves in Fig. 3 show the E>300 keV part of the folded spectra.

Assuming 4% conversion (hν→pe) in a large scale detection device, the above L values correspond to the scintillation output values (Fig. 2) of: S~25, 42, 63 and 100% (S=100% =12000 hν/MeV). The typical best previous In-LS attempt is L=123 pe/MeV. According to the results in Fig. 2, L=300 pe/MeV is attainable with ~9% In in PC and ~15% in MN and L=500 pe/MeV, with ~5% In in MN.

The solid curves in Fig. 3 show the integrated BS rates, i.e. the probability P(BS)/In decay of a correlated BS photon triggering a threshold E, e.g., the lower discriminator level (LDL) on $\gamma_3$. Setting the LDL at 1σ below the $\gamma_3$ energy of 497 keV, LDL(300) = 455 and LDL(500)= 465 keV, for a tag efficiency ε = 0.86. The BS false coincidence rate is:

$R_2(BS) = [(10^6/\eta)] 10^{-5} [10^6 P(BS)/\eta] \eta/s = 5.6 \times 10^{-6}/s$ (4)
$\eta_{min}(BS) = 1.8 \times 10^{12} P(BS)/$ In decay (5)

With P(BS) from Fig. 3 for L(300) and L(500), P(300:455) = $2 \times 10^{-9}$ and P(BS)(500:465) $< 10^{-9}$ and eq. (5),

$\eta_{min}(BS) \sim 3.6 \times 10^3$ to $< 1.8 \times 10^3$ (6),

of the same order as η(4) (eq. 4). The R(BS) background can be reduced further by more than x10 if necessary, by raising the β residue ΔT up to 100 keV in the microcell, at the cost of some tag efficiency. The P(BS) values in fig. 3 display the key achievement of the new In-LS, suppression of the BS background by x100 relative to the old In-LS (the top curve in Fig. 3 for 123 pe/MeV).

*Correlated chance coincidences-Radioimpurities:* Real (βγ) cascades from radioimpurities in the In (the LS is much purer) can also mimic $(e/\gamma)_2+\gamma_3$ in a similar way as R(BS) provided the γ part meets the energy and *spatial* conditions T(e-g), i.e. full containment in the macrocell. Even single γ's could mimic $(e/\gamma)_2+\gamma_3$ via a terminal low energy photoeffect on In (to mimic $(e/\gamma)_2$) with a high probability ~ 1. Energetic γ's from some impurities (e.g. $^{40}$K (1.4 MeV) and from external sources (e.g., 2.6, 1.7 MeV from U/Th) lie well outside the T(f,g) energy gate 450-700 keV and violate the containment rule (T(e)). They can be rejected by inspection of only the candidate events in an *expanded* macrocell.

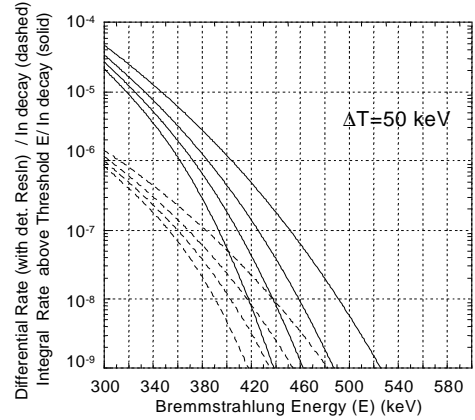

Fig. 3 In-β decay bremmstrahlung (ΔT=50 keV) differential probability/In decay (dashed curves) and integrated intensity above threshold E/In decay (solid), for 4 detector luminosities: (bottom to top) 500, 300, 200 and 123 pe/MeV.

The background problem can now be expressed in a unified way assuming a definite design granularity $\eta=10^5$ and a generalized eq. (5) with P(x) defined as the ratio of the rate of background type x to the β rate of the In decay:

$\eta = 10^5 = 1.8 \times 10^{12} \Sigma P(x); \quad \Sigma P(x) = 55$ (in units of $10^{-9}$) (7).

Eq. (7) sets the background budget for η = $10^5$. So far, we have, P(4) = 2.3 (from the ratio of η(eq. 3) to η=$10^5$) and P(BS) =2. With known (βγ) and γ- branchings of the U/Th/K activities, the specific impurity (βγ) backgrounds are: P(U) ≤ 24, P(Th) ~3.2 and P($K_\gamma$) ~1.2 (~90% of the K γ's rejected by T(e,f,g)). These P(x) correspond to impurities in g/gIn: U≤$10^{-12}$ (mostly from $^{214}$Pb in the U decay chain); Th ~$10^{-12}$ (mostly from $^{228}$Ac), K ~$10^{-9}$. The $^{214}$Pb follows 1600 y Ra in the chain, so that *Ra (and Rn) rather than U* decide the largest background above. This can be precisely monitored *in vivo* by delayed (βα) tags in the succeeding Bi-Po decays in chains. The internal sources thus contribute $\Sigma P(x) \leq 33$, leaving ~22 units P(Ext) for external γ's that satisfy T(e,f,g).

*Neutron Activation of In:* In has high neutron activation cross sections for $^{114}$In* (τ= 70d) at the surface and $^{116}$In* (τ=80min) underground, both with only high energy γ's (~1.3 to 2.8 MeV). At sea level, the saturation activity of $^{114}$In* is 0.5 decays/s/4t In with ~1% γ branching that could contribute a small P(In*). *In vivo* production of $^{116}$In* is higher, ~150 decays/s/4t In. Relatively routine neutron shielding can suppress the $^{116}$In production by a factor ~ $10^4$.

*Cosmogenic Activation:* No isomer cascades fitting the template of the tag T occur in the isotopic vicinity A=90 to 130 except systematic 11/2⁻ isomers. The only important case occurs in $^{115}$Sn itself (at 713 keV, see Fig. 2). This state and the 613 keV state of the In tag itself, can be activated by cosmic ray induced p,α: $^{115}$In(p,n) and $^{115}$In(α,4n) $^{115}$Sb (EC

45min)→$^{115}$Sn (note that the α's from U/Th are below the (α,4n) threshold). The estimated rate[12] of In(pn) is <25/yr/4t In even at the *shallow depth of 800 mwe* with a muon flux ~150/m$^2$h. In any case, these events can be tagged out efficiently, rapidly and locally by the delayed coincidence of n-capture γ-rays in the In with its high σ(n,γ).

The character of background in the indium detector thus presents a dramatically different picture now. The largest background is likely from *non-In* sources P(non-In Int.+Ext) ≤ 52 vs P(4+BS) ~4. That the In decay is no longer the overwhelming issue is due to the new In-LS that reduces R(BS) of the 1985 estimates by some 2 orders of magnitude.

*Detector Design:* The design goal of η=10$^5$ can be achieved in a straightforward way with longitudinal (square section) modules (5x5x 250cm) with 13400 2" phototubes at both ends. With the scintillation luminosities achieved in Fig. 2, ~50 keV events can be located within ~15cm by light time of flight, i.e. in a 375 cc microcell. This implies, with 10 wt% In loading, η ~10$^5$.

A significant reduction in the number of phototubes could be achieved by a simplified version of the "neutrino chamber"[13] first proposed in 1981 (for η~10$^9$). The chamber design combines in addition, efficient calorimetry (minimum inactive material in the detector), a key relaxation of the demand for high transmission lengths in the In-LS and a granularity of η~10$^5$. The principle is illustrated in Fig. 4.

Commercial plastic scintillator light guide (SLG) bars (2-3 mm thick, enclosed in a thin (50-100μ) transparent film that creates an air gap) are immersed in a cross-wise array in a tank of In-LS. The only non-active material in the detector is the film. The SLG material (polystyrene or acrylic) purity can be x10 worse than the In-LS above. The LS contains only the primary scintillation fluor that emits at ~380nm. The secondary shifter which emits at ~430 nm, is *not* dissolved in the LS as usual. It is present only in the SLG. The primary light is absorbed heavily (in a few mm) in the SLG. The reemitted, wave-shifted signal light is conveyed by total internal reflection to the ends of the SLG's with the *same* ~45% light confining efficiency as in a longitudinal square module regardless of cross-sectional dimensions.

The signal output area of the SLG's is much smaller, possibly by x10, than that in the modular design. With suitable multiplexing of the SLG's, the number of phototubes can be reduced by say, a factor 4 to ~3500. Such economy in the electronic channel count could make ≥10 ton In target masses practical. Further, the signal light transmission is determined entirely by the optical properties of the SLG, *not*, that of the bulk In-LS liquid as in the modular design. Thus, the bulk LS light transmission, normally a major design problem for metal-LS, is decoupled from the design criteria. The In-LS needs to be optimized only for the scintillation output and the SLG's independently for optical clarity.

*Neutrino Calibration:* The weak matrix element B(GT) of the In reaction was first theoretically estimated[1] and later confirmed by the (p,n) reaction method to be B(GT)(In 0.613) = 0.17.[2] The B(GT) can be directly calibrated by ν$_e$ activation by a $^{51}$Cr source that emits ν$_e$ lines at 751 and 426 keV. Indeed, the choice of Cr for ν$_e$ calibration was first proposed in the context of In[14]. A 2MCi $^{51}$Cr source in close geometry yields ~1000 751 keV ν$_e$ events in 4mo (=τ ($^{51}$Cr)) in a 4t In detector.

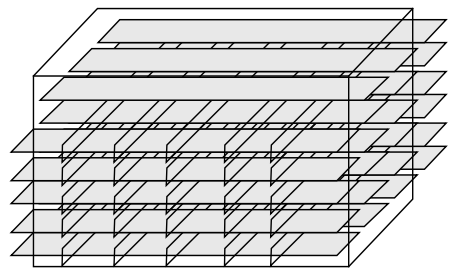

Fig. 4 Principle of the neutrino chamber

In conclusion, based on the new In-LS, the In solar ν$_e$ experiment is technically feasible with broad safety margins. The purity requirement of the target material is non-critical and the depth needed in the underground site, moderate. The target mass for a meaningful experiment is modest and economical (the price of In is ~$100/kg). With the deepest access yet, to the pp ν$_e$ spectrum, the scientific payoff, on the other hand, is unique.

I thank E. Chandross for advice on the LS chemistry T. Kovacs and P. Raghavan for help in the computations, and M. Cribier for helpful critique of the MS and for encouragement.

[1] R. S. Raghavan, Phys Rev Lett 37, 259 (1976)
[2] J. Rapaport et al, Phys. Rev. Lett. 54, 2325 (1985)
[3] R. S. Raghavan, Science 267, 45 (1995).
[4] P. Krastev and S. Petcov , Phys Rev D53, 1665 (1995).
[5] J. N. Bahcall et al, hep-ph/0103179 (March 2001).
[6] L. Pfeiffer et al, Phys Rev Lett 41, 63 (1979); Y. Suzuki et al, NIM A239, 615 (1990).
[7] R. S. Raghavan, Phys Rev Lett 78, 3618 (1997).
[8] Documentation on the LENS project at the Laboratori Nazionali del Gran Sasso (Letter of Intent (1999), Progress Reports (2000, 2001) are available from masciulli@lngs.infn.it
[9] R. S. Raghavan, *Proc.Conf.Status &Future of Solar Neutrino Research,* Brookhaven Natl. Lab. BNL-50879 (1978) vol.II, p1.
[10] M. Bourdinaud et al, DAPNIA/Saclay report (1984).
[11] M. Cribier et al, DAPNIA/Saclay report (1984). This calculation of the external BS in In metal (Z=49) overestimates by ~x7 that for In LS with 5-10% In (<Z>~7) since the BS intensity ∝ Z. This is still adopted here since it does not include the *internal BS*, lacking an IBS theory for the 4$^{th}$ forbidden β-decay of In. With the known IBS-*allowed* β-decay estimate, the overestimate is probably x2.
[12] The estimates were made relative to $^{37}$Ar production rates at Homestake at shallow depths (thus, mostly $^{37}$Cl(p,n)) (Davis et al, *Solar Neutrinos & Neutrino Astronomy*, Ed. M. Cherry et al, AIP Conf. Series #126 (1985), p.1) and scaling for the number of $^{37}$Cl and $^{115}$In nuclei and the measured σ(pn) $^{115}$In/σ(pn) $^{37}$Cl ~ 0.25.
[13] R. S. Raghavan, *Neutrino '81*, Ed. R. Cence (1981), vol I, p. 27
[14] R. S. Raghavan, BNL-50879 (ref. 6) (1978), vol. II, p.270.